\documentclass[aps,prl,twocolumn,groupedaddress,showpacs]{revtex4}

\usepackage[bookmarks]{hyperref}
\usepackage[dvips]{graphicx}
\usepackage{amsmath}
\usepackage{amstext}
\usepackage{graphics}
\usepackage{exscale}
\usepackage{epsfig}
\usepackage[T1]{fontenc}
\usepackage{bm}
\usepackage{bbm}
\usepackage{color}
\usepackage[normalem]{ulem}

\usepackage{version}

\usepackage{latexsym}

\usepackage[tight]{subfigure}

\renewcommand{\epsilon}{\varepsilon}

\newcommand{\elcre}[2]{ c^{\dagger}_{#1,#2}}
\newcommand{\elann}[2]{ c^{\phantom{\dagger}}_{#1,#2}}

\newcommand{\e}{\mathrm e}

\newcommand{\vct}[1]{\bm #1}
\newcommand{\vk}{{\bm k}}
\newcommand{\vq}{{\bm q}}

\newcommand{\CC}{\mathcal{C}}
\newcommand{\GG}{\mathcal{G}}

\includeversion{commentst1} 
\newcommand{\eqfigscl}[2]{\vcenter{\hbox{\includegraphics[scale=#1]{#2}}}}
\newcommand{\eqfigsclbot}[2]{{\hbox{\includegraphics[scale=#1]{#2}}}}

\begin{document}

\title{Dynamical instabilities and transient short-range order in the fermionic Hubbard model}

\author{Johannes Bauer,${}^1$ Mehrtash Babadi,${}^2$ and Eugene Demler${}^1$}
%\author{Johannes Bauer${}^1$ and Mehrtash Babadi${}^2$}
\affiliation{${}^1$Department of Physics, Harvard University, Cambridge,
  Massachusetts 02138, USA}
\affiliation{${}^2$Institute for Quantum Information and Matter, Caltech, Pasadena, CA 91125}
\date{\today} 

\begin{abstract}
We study the dynamics of magnetic correlations in the half-filled fermionic
Hubbard model following a fast ramp of the repulsive 
interaction. We use Schwinger-Keldysh self-consistent second-order
perturbation theory to investigate the evolution of single-particle Green's
functions and solve the non-equilibrium Bethe-Salpeter equation to study the
dynamics of magnetic correlations.  This approach gives us new insights into
the interplay between single-particle relaxation dynamics and the growth of
antiferromagnetic correlations. Depending on the ramping time and the final
value of the interaction, we find different dynamical behavior 
which we illustrate using a dynamical phase diagram. Of
particular interest is the emergence of a transient short-range ordered regime
characterized by the strong initial growth of antiferromagnetic correlations  
followed by a decay of correlations upon thermalization. The discussed
phenomena can be probed in experiments with ultracold atoms in optical
lattices.

\end{abstract}
%71.30.+h
%Metal-insulator transitions and other electronic transitions 
%71.27.+a
%Strongly correlated electron systems; heavy fermions 
%71.10.Fd
%Lattice fermion models (Hubbard model, etc.)
%71.10.Hf
%Non-Fermi-liquid ground states, electron phase diagrams and phase transitions in model systems 
%75.40.Gb
%Dynamic properties (dynamic susceptibility, spin waves, spin diffusion, dynamic scaling, etc.) 
%64.60.A- 	
%Specific approaches applied to studies of phase transitions
%64.60.Bd 	General theory of phase transitions
\maketitle

Non-equilibrium dynamics of quantum many-body systems has been the subject of 
experimental inquiry in many areas of physics in the recent years. 
For example, pump-probe experiments in solid-state systems have 
addressed such important issues as the observation of the Higgs mode in
superconductors \cite{MHMUTWS13,MTFSMUTWAS14} and the identification of dominant couplings in
cuprate superconductors \cite{Pea07,Dea12,Mea12,Sea12b}. 
A particularly exciting direction is the dynamical generation, suppression, or
manipulation of ordered phases using external fields. Non-equilibrium induced
superconductivity~\cite{Fea11,Hea13} and ultrafast melting of
charge-density-wave order~\cite{TPBC08,Sea08a,SGLKDF13} in cuprate
superconductors, transient generation of spin-density-wave order in
pnictides~\cite{Kea12b}, and ultrafast manipulation of the order in
multiferroics~\cite{Kea14} are examples of such possibilities. 
\begin{figure}[!thpb]
\centering
\subfiguretopcaptrue
\includegraphics[width=0.46\textwidth]{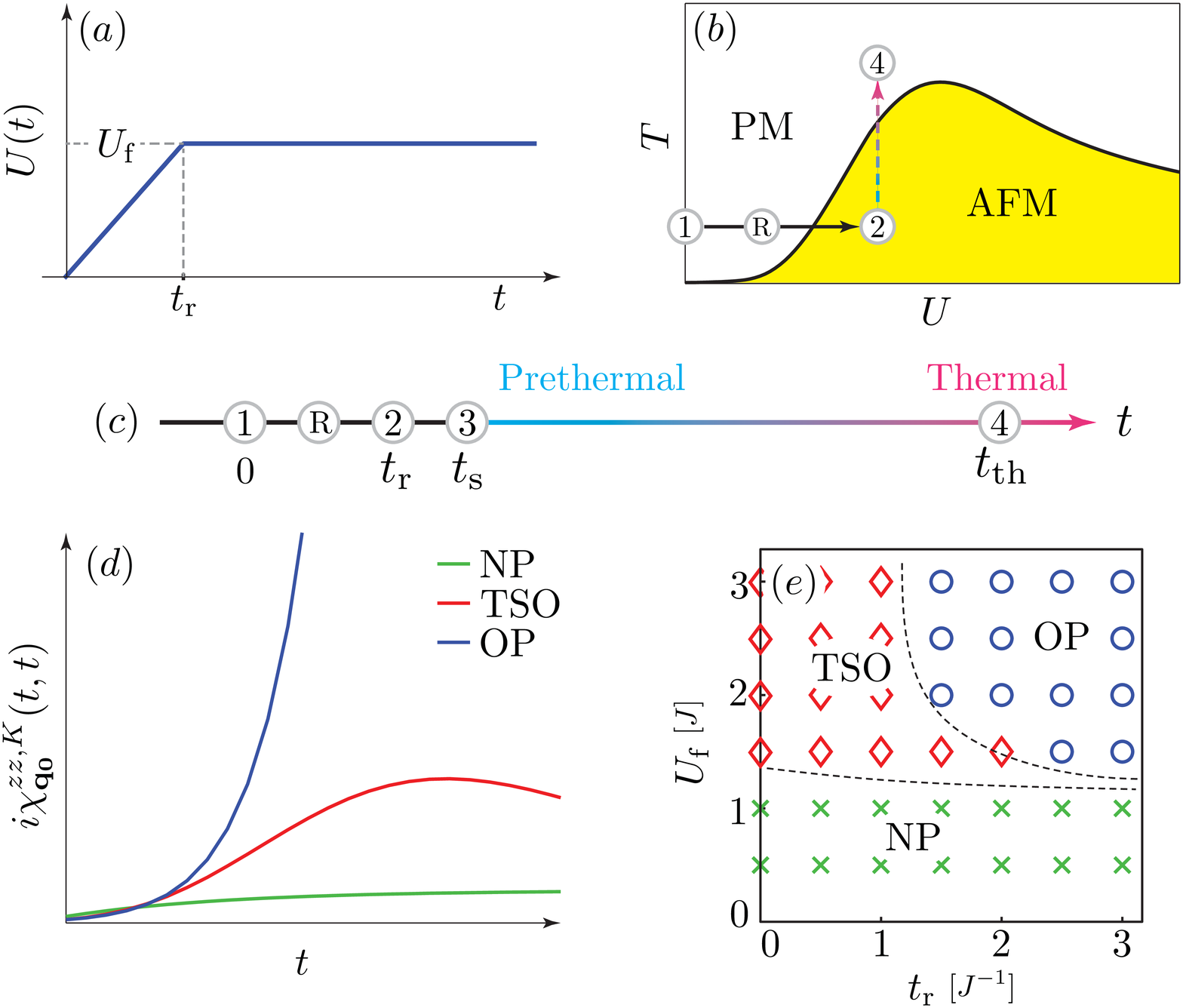}
\caption{(color online). (a) Linear interaction ramp $U(t)$ with ramping time
  $t_{\rm r}$ and final interaction $U_{\rm f}$, (b) Schematic equilibrium
  phase diagram showing the paramagnetic phase (PM), the critical
  temperature $T_c^{\rm eq}(U)$ for the antiferromagnetic (AFM) phase, and
  the relaxation trajectory to a final temperature $T_{\rm f}$ (c) Time scales and
  different regimes (see text), (d) Qualitatively different regimes for the evolution of
  the equal-time AFM correlations; NP: slow growth to normal phase result.
  TSO: transient short-range order; AFM correlations initially develop and later decay upon thermalization,
   $T_{\rm f}>T_c^{\rm eq}(U_{\rm f})$. OP: AFM correlations grow and final
  thermalized state expected to be the ordered phase,
  $T_{\rm f}<T_c^{\rm eq}(U_{\rm f})$. (e) The dynamical phase diagram showing different
  regimes after a ramp from an initial PM state at temperature $T_i=0$ as
  a function of $U_{\rm f}$ and $t_{\rm r}$. The dashed lines are meant as a guide to the eye.}        
\vspace*{-0.6cm}
\label{fig:schematic}
\end{figure}

Artificial systems of ultracold atoms allow a clean and
tunable experimental realization of the paradigmatic condensed matter models that underlie
many solid-state systems. In these experiments, microscopic parameters can be rapidly changed 
using external fields and non-equilibrium quantum dynamics can be
probed~\cite{GMHB02,Sea08c,CGSD08,BDZ08,Bea09,PSSV11,Cea12b,Sea12,Fea14preb,Tea14pre}. 
For example, in Ref.~\cite{Bea09} Jo {\em et al.} reported an experimental study of the
possible occurrence of the Stoner ferromagnetic instability following a rapid
interaction quench to the BEC side of a Feshbach resonance with large positive
scattering length (for subsequent
experiments and analysis see Refs.~\cite{Pea11,Sea12}).  

Here, we study dynamical instabilities and the growth of magnetic
correlations in the repulsive fermionic Hubbard model following an interaction ramp.
At half-filling, the paramagnetic (PM) state is unstable toward
antiferromagnetic (AFM) ordering for weak on-site repulsion at low
temperatures [see Fig.~\ref{fig:schematic}~(a,b)].
One of the central findings of our study is the identification of an extended
parameter regime in which the prethermal state that emerges after the
interaction ramp~\cite{BBW04a,MK08} exhibits growing AFM correlations, and can
develop sizable domains with short-range AFM order; interestingly, these features
are only transient and decay when the thermal equilibrium state is
approached [see Fig.~\ref{fig:schematic}~(d,e)]. Phenomenologically,
``transient short-range order'' (TSO) in the present context can be understood
by first noting that at half filling, the logarithmic divergence of the spin
susceptibility that results from Fermi surface nesting is only suppressed by
finite temperature. The prethermal single-particle momentum distribution
$n_{\vk}^\mathrm{pt}$ is found to closely resemble the initial low-temperature
distribution $n_{\vk}^0$ for $k \approx 
k_\mathrm{F}$~\cite{MK08} and thus elicits a strong
AFM response and even an instability for large enough on-site
repulsion. The instability is maintained for a time
inversely proportional to the thermalization rate of the low-energy prethermal
quasiparticles. The disordered PM state is eventually recovered as $n_\vk(t)$
slowly approaches the final thermal state in which the 
generated temperature $T_{\rm f}$ exceeds $T_c^{\rm eq}$, the critical AFM
transition temperature. 

Considerable progress has been made in the accurate description of many-body
quantum dynamics in one dimension~\cite{Caz06,MWNM07,KLA07,RDO07,RDO08,Rig09,DHG11} and in infinite  
dimensions using the non-equilibrium extension of dynamical mean 
field theory (DMFT)~\cite{ATEKOW14}. The situation is more
challenging in two and three dimensions where accurate
and efficient methods as such are not available. Previous works on the
non-equilibrium dynamics of the Hubbard model~\cite{MK08,EKW09,MK10,SM10,SK13pre} and itinerant 
fermions~\cite{MSF14pre} have studied thermalization following interaction
ramps, which is found to be preceded by the rapid establishment of a
prethermal plateau with a substantially modified $n_\vk$. 
On another front, the dynamics of the order parameter in quenches has been studied
within the integrable BCS theory~\cite{BLS04,BL06,YTA06}. These works,
however, do not take into account single-particle excitations and order
parameter fluctuations at finite momentum, both of which break
integrability and substantially modify the physics in low dimensions.
More recently, quenches from the ordered AFM phase into the normal phase have been analyzed
within DMFT \cite{WTE12,TEW13,SF13}, as well as slow ramps into the AFM state
starting with a small seeding field~\cite{TW13}. To our knowledge,
none of these works have addressed the interplay between
single-particle dynamics and the collective modes
during the relaxation dynamics. 
As briefly described earlier, we show that this interplay introduces
additional complexity and richness to the non-equilibrium dynamics.

\paragraph{Model and formalism -}

We consider the quasi-two-dimensional Hubbard model \cite{fnot1}
with nearest neighbor hopping and a time-dependent on-site interaction: 
\begin{equation}
H=\sum_{\vk,{\sigma}}\epsilon_{\vk} \, \elcre \vk{\sigma} \elann
\vk{\sigma}+U(t)\sum_i n_{i,\uparrow}n_{i,\downarrow}. 
\label{hubham}
\end{equation}
The dispersion is $\epsilon_{\vk} = -2J(\cos k_x + \cos k_y)$, where $J$
is the nearest neighbor hopping amplitude. We work in the units
where $\hbar = k_{\rm B} = J =1$ and assume half-filling $\langle n_{\uparrow}
\rangle = \langle n_{\downarrow} \rangle = 1/2$ hereafter.
The dispersion satisfies the perfect nesting condition
$\epsilon_{\vk+\vq_0}=-\epsilon_{\vk}$ 
for $\vq_0=(\pi,\pi)$, and the PM state exhibits an AFM instability signaled
by the divergence of 
the static magnetic susceptibility $\chi^{zz}_{\vq_0}(i\omega=0)$. At weak coupling, the critical
temperature $T_c$ can be estimated within RPA, $T_c\sim
J \e^{-\sqrt{c J/U}}$ where $c$ is a numerical
constant~\cite{Don91}. 
Higher order correction analogous to the ones discussed by Gor'kov~\cite{GM61}
in the theory of superconductivity are found to be important
already at weak coupling and result in an $\mathcal{O}(1)$ correction to the
prefactor of $T_c$; this correction can be conveniently captured by replacing
$U\to U_{\rm eff}$ in the RPA calculation~\cite{Don91,MF92,HPSV00,TBCC05,supp}.  
A first estimate of the growth rate of the staggered magnetization in the
PM state, $\Delta_{\vq_0}$, can be obtained from linear response~\cite{Pea11},
and one finds $\Delta_{\vq_0}\sim J \e^{-\sqrt{\bar c J/U_{\rm
      eff}}}$. Nonlinear corrections quickly become relevant due to the fast
single-particle relaxation dynamics, making this result questionable for
longer times.

Going beyond linear response, we describe the dynamics within the framework of
$\Phi$-derivable approximations \cite{BK61} and non-equilibrium Green's
functions on the Schwinger-Keldysh contour~\cite{Kel65,Ram07,Kam11}. The 
closed-time-path single-particle Green's function is defined as  
$\GG_{i,\sigma;j,\sigma'}(t, t') = -i \big\langle T_\CC[c^{\phantom{\dagger}}_{i,\sigma}(t_1)\,c^\dagger_{j,\sigma'}(t_2)]
\big\rangle$, where $\CC$ is the round-trip Schwinger-Keldysh time contour,
$t_{1}, t_2 \in \CC$ are contour times and $T_\CC$ is the contour
time-ordering operator. The initial state of the system at $t=0$ is assumed to
be a uniform and uncorrelated paramagnet. In this setup, the
$SU(2)$ and translation symmetry is preserved at all times such that $\GG_{i,\sigma;j,\sigma'} \equiv
\delta_{\sigma\sigma'} G_{i-j}(t,t')$. Dynamical symmetry breaking requires a
weak inhomogeneity or small seeding field, which we do not assume here; 
rather we probe the growth of magnetic correlations by studying the non-equilibrium
spin-spin correlation function $\chi^{\rm K}_{\vq}$, and the growth (instability) of
domains from the retarded response $\chi^{\rm R}_{\vq}$. 

In the momentum basis, $G_\vq(t,t')$ is obtained by solving the non-equilibrium Dyson
equation $G^{-1}_\vq(t,t') = G_{0,\vq}^{-1}(t,t') - \Sigma_\vq(t,t')$. Here,
$G_{0,\vq}^{-1}(t,t') = (i\partial_t - \xi_{\vq})\,\delta_\CC(t,t')$, where
$\delta_\CC$ is the contour $\delta$-function, and the self-energy is obtained
as $\Sigma_\vq(t,t') = -\delta \Phi[\GG]/\delta 
G_{-\vq}(t',t)$, where $\Phi[\GG]$ is the Luttinger-Ward functional. We
consider ramps to weak final interactions $U_{\rm f}<4J$ such that a
skeleton expansion of $\Phi[\GG]$ up to the second order in $U$ is justified: 
\begin{equation}
\includegraphics[scale=1.65]{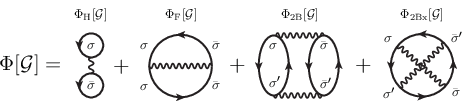}.
\end{equation}
\noindent These vacuum diagrams determine $\Sigma$ and the irreducible vertex
$I(11';22') = \delta^2 \Phi[\GG]/ \delta \GG(1'1) \delta \GG(22')$ in the
particle-hole channel. The latter is used to calculate the spin-spin
correlation function $\chi^{\mu\nu}_\vq(t,t') 
\equiv -i\big\langle T_\CC[\hat{S}^\mu_{\vq}(t) \,
\hat{S}^\nu_{-\vq}(t')\big\rangle$ by solving a non-equilibrium
Bethe-Salpeter equation. Here $\hat S^{\mu}(\vq) = \frac{1}{2}\sum_{\vk}c^\dagger_{\vk +
  \vq,\alpha}\,\sigma^{\mu}_{\alpha\beta} \,c^{\phantom{\dagger}}_{\vk,\beta}$
with $\{\sigma^\mu\}$ being the Pauli matrices. The $SU(2)$ symmetry implies
$\chi_\vq^{\mu\nu}(t,t') = \frac{1}{2}\,\delta^{\mu\nu} 
\chi_\vq^{+-}(t,t')$, where $\chi^{+-}$ is the transverse spin-spin
correlator and its diagrammatic calculation is more economical than the diagonal correlators.

Carrying out such calculations in real-time and on a dense two-dimensional
momentum grid is numerically extremely challenging, even for low order $\Phi$-derivable approximations.
We therefore make additional simplifying approximations to
proceed. First, we approximate the irreducible vertices by their local 
parts, i.e. $\Sigma_\vq(t,t') \to \Sigma_\ell(t,t')$, where
$\Sigma_\ell(t,t')$ is a $\vq$-independent self-energy similar
as in DMFT. The local approximation captures the full temporal structure
of the vertices while significantly simplifying the forthcoming
analysis. Also, the momentum dependence of $\Sigma_\vq$ is
known to be fairly weak at weak-coupling \cite{ZSS95,NM03,supp}. The
$SU(2)$ symmetry of the state implies $\Sigma_{\rm F} = \Sigma_{\rm 2Bx} = 0$, and we find:
\begin{align}
\Sigma_\ell(t,t') &= \eqfigsclbot{2}{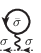} \,\,\, + \,\,\, \eqfigsclbot{2}{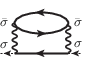} = U(t)\, n \, \delta_\CC(t,t')\nonumber\\
&- U(t)\,U(t') \, G_\ell(t,t') \, \Pi_\ell^{\rm ph}(t,t'),
\end{align}
where $n=1/2$ is the filling, $G_\ell(t,t') = \frac{1}{N}\sum_\vq G_\vq(t,t')$
is the local Green's function, and $\Pi_\ell^{\rm ph}(t,t') = G_\ell(t,t') \,
G_\ell(t',t)$. The Hartree term only gives a dynamical phase and can be
gauged out using the particle-hole symmetry of the half-filled state.
The second-order self-energy, however, is non-trivial and describes the single-particle
relaxation dynamics. The transverse spin correlator in the framework of
$\Phi$-derivable 
approximations is obtained by supplementing the real-time action with a fictitious
transverse magnetic field term $-\int_\CC \mathrm{d} t \sum_{\vq}
B_\vq(t)\,\hat{S}^-_{-\vq}(t)$ and calculating the induced linear variation in the
Green's function $\chi^{+-}_\vq(t_1,t_2; t') \equiv
\delta\,\mathrm{Tr}[\GG_\vq(t_1,t_2; B) \,S^+] / \delta B_\vq(t')$. The result
is a contour Bethe-Salpeter equation (BSE): 
\begin{align}\label{eq:BSE}
&\chi^{+-}_\vq(t_1,t_2;t') = \Pi_\vq^{\rm ph}(t_1,t_2;t'^+,t') + \int_\CC \mathrm{d} t'_1\nonumber\\
&\times \int_\CC \mathrm{d} t'_2 \, \Pi_\vq^{\rm ph}(t_1,t_2;t'_1,t'_2) \, I_\ell(t'_1,t'_2) \, \chi^{+-}_\vq(t'_2,t'_1;t'),
\end{align}
where $\Pi_\vq^{\rm ph}(t_1,t_2;t'_1,t'_2) = \frac{1}{N}\sum_{\vk}
G_{\vk+\vq}(t_1,t'_1) \, G_{\vk}(t'_2,t_2)$ 
%is the polarization 
and $I_\ell(t_1,t_2) = I_\mathrm{F}(t_1,t_2) + I_\mathrm{2Bx}(t_1,t_2)$: 
\begin{equation}
I_\ell(t_1,t_2) = \eqfigscl{2}{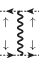} \,\,\, + \,\,\, \eqfigscl{2}{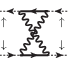},
\end{equation}
is the local irreducible vertex, consisting
of a Fock (ladder) part $I_\mathrm{F}(t_1,t_2) = iU(t_1) \,
\delta_\CC(t_1,t_2)$, and a second-order ``Gor'kov'' part
$I_\mathrm{2Bx}(t_1,t_2) = - U(t_1)\, U(t_2) \, \Pi_\ell^{\rm pp}(t_1,t_2)$,
where $\Pi_\ell^{\rm  pp}(t_1,t_2)=G_\ell(t_1,t_2) \,G_\ell(t_1,t_2)$. These
vertex parts arise from $\Phi_{\rm F}$ and $\Phi_{\rm 2Bx}$ vacuum diagrams,
respectively. Finally, the transverse spin correlator is calculated as
$\chi^{+-}_\vq(t,t') \equiv \chi^{+-}_\vq(t,t^+; t')$. 
  
Even in the local approximation, the numerical solution of Eq.~(\ref{eq:BSE})
for the 3-time function $\chi^{+-}_\vq(t_1,t_2;t')$ is formidable and requires
the inversion of very large matrices. 
Had the vertex $I_{\rm 2Bx}(t'_1,t'_2)$ been local in time (as in
$I_\mathrm{F}$), the BSE in Eq.~\eqref{eq:BSE} could be immediately reduced to
a numerically tractable integral equation for the 2-time correlator $\chi^{+-}_\vq(t,t')$, with only one intermediate
contour integral. 
This motivates us to approximately incorporate the role
of $I_\mathrm{2Bx}$ vertex correction via an effective time-local
vertex. 
For temperatures $T \ll W$ (bandwidth $=8 J$) and near-equilibrium states, 
the spread of $I_\mathrm{2Bx}(t,t')$ on $t - t'$ is of the order of
$W^{-1}$, which is considerably smaller than $\Delta^{-1}_{\vq_0}$, the
inverse growth rate mentioned before.
Therefore, beyond the numerical reduction of the
non-equilibrium instability rate, no qualitatively distinct behavior is
expected to emerge as a matter of the temporal non-locality of the vertex
$I_\mathrm{2Bx}$. As mentioned before, the equilibrium Gor'kov correction can
be obtained by replacing $U \to U_{\rm eff}[U]$ in the RPA
calculation~\cite{Don91,MF92,HPSV00,TBCC05}, 
where $U_\mathrm{eff}[U]$ is found by requiring that the correct AFM transition
temperature is reproduced. Here, we assume that
the same approximate picture holds for the weak-coupling non-equilibrium 
dynamics as well, and use the equilibrium effective interaction as a
time-local vertex, albeit at the instantaneous value of $U(t)$, i.e. 
$I_\ell(t_1,t_2) \rightarrow
iU_\mathrm{eff}[U(t_1)]\,\delta_\CC(t_1,t_2)$. This allows us to set $t_2 = t_1^+$ in Eq.~\eqref{eq:BSE} and simplify it to:
% to an equation for the 2-time spin correlator: 
\begin{align}
\chi^{+-}_\vq(t,t') &= \Pi_\vq^{\rm ph}(t,t')+\nonumber\\
&\int_\CC \mathrm{d} t'' \, \Pi_\vq^{\rm ph}(t,t'')\,iU_\mathrm{eff}[U(t'')] \, \chi^{+-}_\vq(t'',t'),
\end{align}
where $\Pi_\vq^{\rm ph}(t,t') = \Pi_\vq^{\rm ph}(t,t^+;t'^+,t')$.
For the numerical solution method of the non-equilibrium Dyson equation and the above BSE, see \cite{supp}.

\begin{figure}[!t]
%\vspace*{1cm}
\centering
\includegraphics[width=0.483\textwidth]{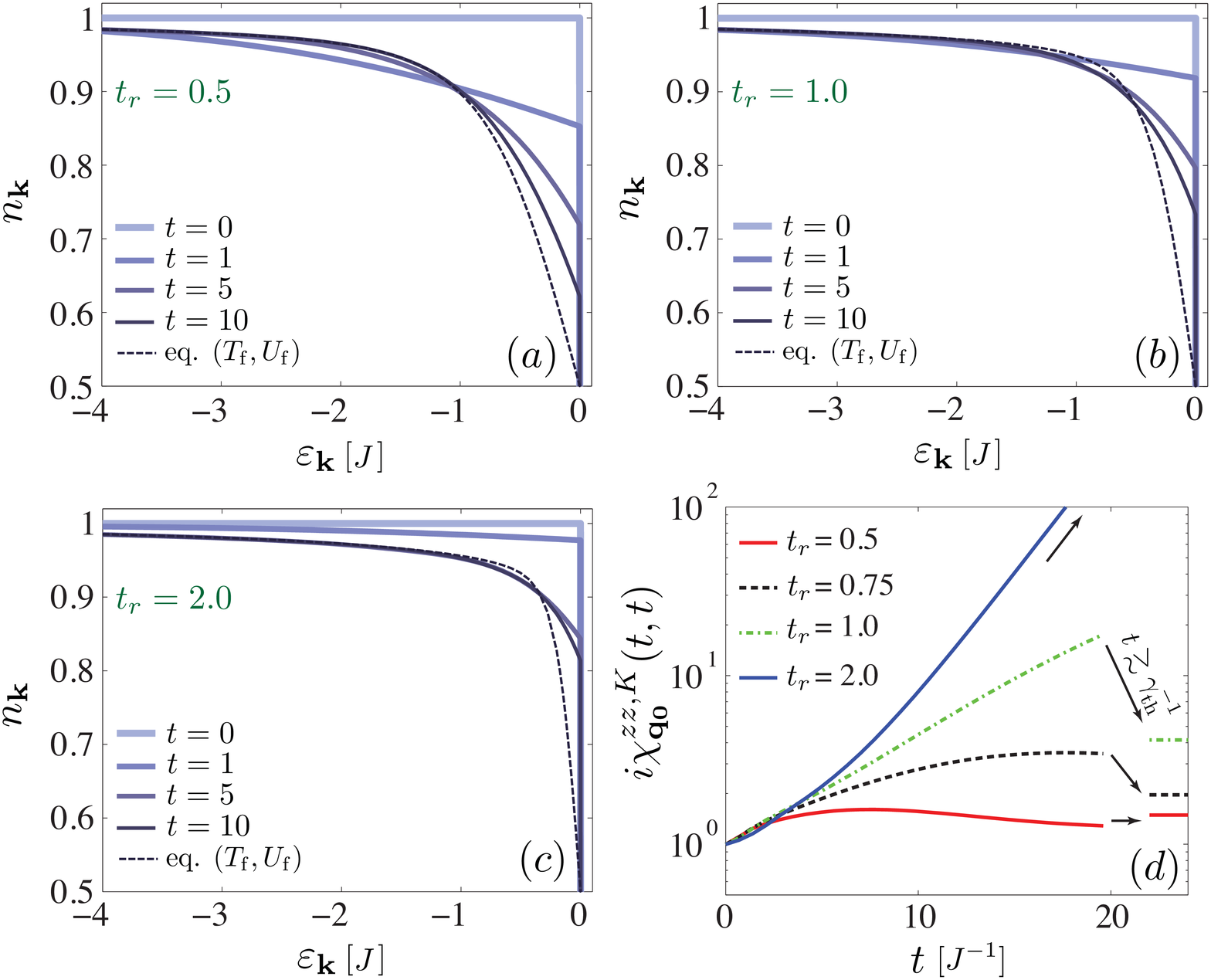}
\vspace*{-0.5cm}
\caption{(color online). Momentum distribution function
  $n_{\vk}(t)$ for $T_{\rm i}=0$, $U_{\rm f}=3$ and different ramping
  times. (a) $t_{\rm r} = 0.5$, $T_{\rm  f}\approx 0.29  >
  T_c^{\rm eq}$, (b) $t_{\rm r} = 1.5$, $T_{\rm f} \approx 0.15  >
  T_c^{\rm eq}$, (c) $t_{\rm r}=2$, $T_{\rm f}\approx 0.08  <
  T_c^{\rm eq}$. The dotted lines show the final equilibrium $n_{\vk}$ at
  $T=T_{\rm f}$. (d) the time dependence of $\chi^{zz,K}_{\vq_0}(t,t)$ for
  different $t_{\rm r}$ in TSO and OP regime (semi-log plot). The long-time limit of
  the correlators in the TSO regime is shown on the plot from an equilibrium
  calculation at $T=T_{\rm f}$ \cite{fnot2}.} 
\vspace*{-0.5cm}     
\label{fig:nkchiK}
\end{figure}

\paragraph{Results and Discussion -} 
For concreteness, we consider an uncorrelated PM
state at initial temperature $T_{\rm i}=0$ subject to a linear interaction
ramp to a final value of $U_{\rm f}$ within a time interval $t_{\rm r}$
[Fig.~\ref{fig:schematic}(a)]. The timeline of the single-particle dynamics is shown schematically in
Fig.~\ref{fig:schematic}(c). Following the ramp, a brief switching
regime with a duration $t_s \sim 1/J$ is observed~\cite{MK10} which leads to a prethermal single-particle momentum
distribution $n_{\vk}^{\rm pt}$ that deviates from the initial distribution by
$\mathcal{O}(U_{\rm f}^2)$~\cite{MK08}. Collisions slowly smear 
$n_{\vk}^{\rm pt}$ to a thermal distribution (see Fig.~\ref{fig:nkchiK}). The
thermalization rate of the low-energy quasiparticles
% $\epsilon \simeq \epsilon_F$ 
is found as $\gamma_{\rm th}
\sim U_{\rm f}^4 /J^3$ for short ramps, and a smaller value $\gamma_{\rm th} \sim U_{\rm f}^4/(J^5 t_{\rm r}^2)$ for long ramps~\cite{MK10}.
The final temperature $T_{\rm f}$ generically increases with $U_{\rm f}$ and decreases with $t_{\rm r}$. We monitor the evolution of the single-particle momentum distribution
$n_{\vk}(t)$, and the equal-time Keldysh correlator $\chi^{\rm K}_\vq(t,t)$$=$$-i\langle \{\hat{S}^+_\vq(t),
\hat{S}^-_{-\vq}(t)\}\rangle$, and the retarded spin correlator
$\chi^{\rm R}_\vq(t,t') = -i\theta(t-t')\langle 
[\hat{S}^+_\vq(t), \hat{S}^-_{-\vq}(t')] \rangle$.

We identify qualitatively different behaviors depending on $U_{\rm f}$ and $t_{\rm
  r}$, which is concisely collected in the dynamical phase diagram shown in
Fig.~\ref{fig:schematic}(e). The symbols (NP, $\times$) correspond to weak
grow to normal phase, (TSO, $\diamond$) to transient AFM 
correlations along with a PM state upon thermalization, and (OP, $\circ$) to AFM ordered phase upon thermalization. 
\noindent
The evolution of the AFM correlations for each of these dynamical modalities
is shown schematically in Fig.~\ref{fig:schematic}(d). The NP regime is
identified by a monotonic growth of the equal-time spin correlator
$\chi_{\vq}^{\rm K}$ to its final equilibrium value and being bounded by it, along
with a decaying spin response ($\chi_{\vq}^{\rm R}$, see~\cite{supp}) for all
modes. In the TSO regime, one observes an enhancement of AFM correlations for
intermediate times $t_{\rm s} \lesssim t \lesssim \gamma_{\rm th}^{-1}$;
in this regime, AFM seeds can rapidly grow into sizeable domains as signaled
by the exponentially growing retarded response function. These
features eventually subside at longer times $t \gtrsim \gamma_{\rm th}^{-1}$
as the system thermalizes in the disordered PM state. Finally, in the OP case,
AFM correlations keep growing 
exponentially and the final thermal state is expected to be ordered. The 
detailed long time evolution in this state depends on inhomogeneities present
in any real system and requires a fully self-consistent treatment of the
emerging order parameter, which is beyond the scope of this paper.

Fig.~\ref{fig:nkchiK} shows examples of the evolution of the instantaneous
momentum distribution $n_{\vk}(t)=\frac{1}{2} - \frac{i}{2}\,G_\vk^{\rm
  K}(t,t)$ and spin-spin correlation function $i\chi^{\rm
  K}_{\vq_0}(t,t)$ for $T_{\rm i}=0$ and $U_{\rm f}=3$ in the TSO regime
($t_{\rm r}=0.5,\,1$) and the OP regime ($t_{\rm  r}=2$).  
As discussed before, $\gamma_{\rm th}$ and $T_{\rm f}$ decrease with increasing
$t_{\rm r}$, such that prethermal regimes are maintained for longer times.
This allows the AFM correlations in the TSO regime to grow to
sizeable values, as seen in Fig.~\ref{fig:nkchiK}(d). Since $T_{\rm
  f}>T_c^{\rm eq}$, $i\chi^{K}_{\vq_0}(t,t)$ is eventually expected to subside 
to the thermal equilibrium result in all cases. 
The regime OP is realized in panel (c) where the ramp time $t_{\rm r}=2$ is
longer, the heating is lower, and the system can thermalize in an ordered
phase. 
\begin{figure}[!t]
%\vspace*{1cm}
\centering
\includegraphics[width=0.48\textwidth]{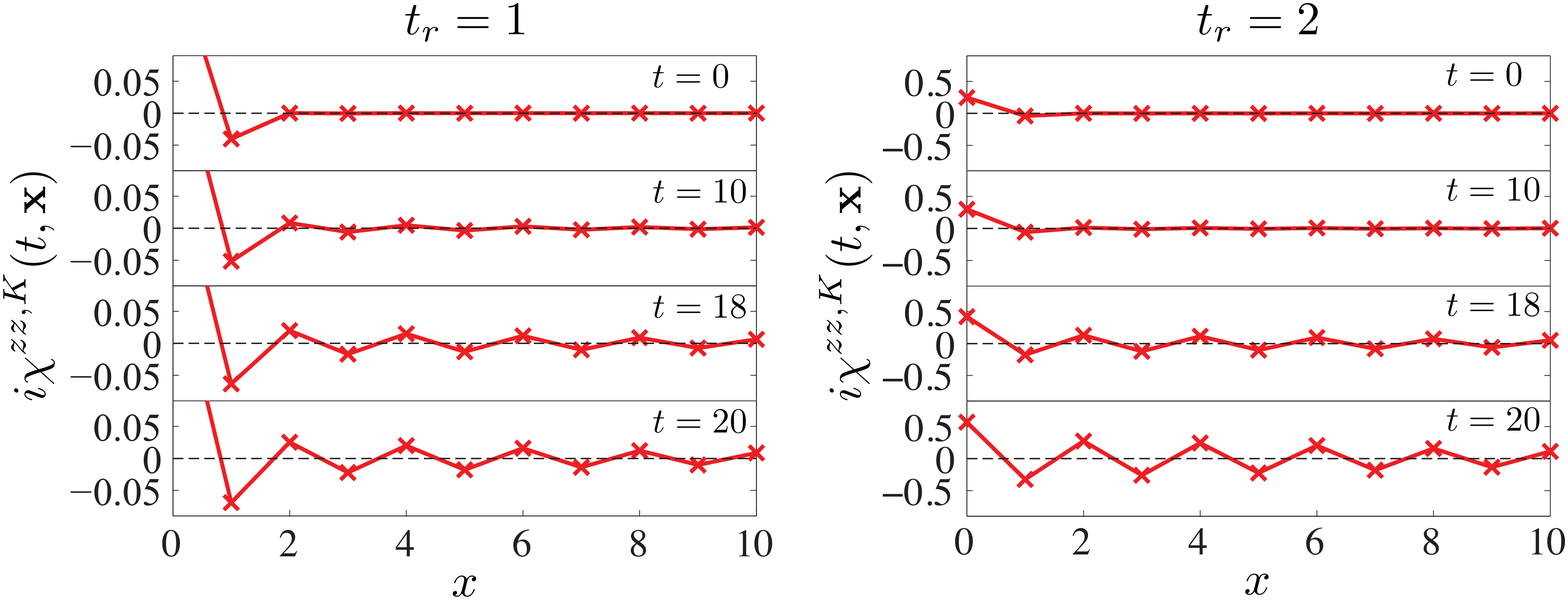}
\vspace*{-0.5cm}
\caption{(color online). Equal time spatial spin correlation function
  $i\chi^{\rm K}(t,x)$ for $t_{\rm r}=1$ (left, TSO), and $t_{\rm r}=2$
  (right, OP) for $T_{\rm i}=0$ and $U_{\rm f}=3$ for different times $t$ and lattice spacing
  $a=1$. Notice the different scale on the vertical axis.} 
\vspace*{-0.3cm}
\label{fig:chiK_r_dift}
\end{figure}
\noindent
Finally, the growth of AFM correlations in real space after the ramp can be
seen by calculating $\chi^{\rm K}(t,\vct r)=\frac{1}{N}\sum_{\vq} \e^{i \vq \vct r}
\,\chi_\vq^{\rm  K}(t,t)$, as shown in Fig.~\ref{fig:chiK_r_dift} for $t_{\rm
  r}=1$ (TSO), $2$ (OP). A clear AFM pattern develops once $i\chi^{\rm K}_{\vq_0}(t,t)$
has grown to large enough values.

\paragraph{Conclusions -}
We have studied the evolution and interplay of fermionic quasiparticles and
collective magnetic correlations in the Hubbard model at half-filling
following an interaction ramp, and have identified three regimes of qualitatively different dynamical
behavior. Of particular interest is the 
occurrence of a parameter regime in which the prethermal state
is marked with strong but transient AFM correlations. 

The non-equilibrium phenomena discussed here can be probed 
in ultracold atoms experiments using
measurements of local spin correlations \cite{GUJTE13},
  Bragg scattering of light \cite{Hea14pre}, time-of-flight and noise correlation
  measurements \cite{FGWMGB05,GRSJ05,RBVSFPB06} once low enough temperatures are
achieved. In fact, a significant enhancement of the AFM correlations has been
reported recently~\cite{Iea14,Hea14pre}. We point out that questions
addressed in this paper are generally important for the many ongoing experimental
efforts for realizing quantum simulators of the fermionic Hubbard
model. Inelastic losses in the vicinity of Feshbach resonances are fast and
the experiments need to be performed rapidly to avoid strong heating
of the atoms. Separating transient dynamical phenomena from equilibrium properties
is crucial for drawing conclusions from such experiments.

Our work further opens the interesting new direction of designing protocols to realize
novel many-body states using metastable prethermal 
states, in particular, states which may not be realized in equilibrium. Finally, our
results show that fermionic systems with gapless excitations can introduce new
features to the Kibble-Zurek picture of domain formation and 
coarsening in the dynamical crossing of phase boundaries discussed in the
context of purely bosonic systems~\cite{ZDZ05,Pol05,CL06}.

\paragraph{Acknowledgments -} We wish to thank E. dalla Torre, M. Knap, A. Millis, J. Han,
M. Schiro, P. Strack for helpful discussions.  JB acknowledges financial support from the DFG
through grant number BA 4371/1-1. MB was supported by the Institute for
Quantum Information and Matter, an NSF Physics Frontiers Center with support
of the Gordon and Betty Moore Foundation We also acknowledge support from
Harvard-MIT CUA, DARPA OLE program, AFOSR Quantum Simulation MURI, AFOSR MURI
on Ultracold Molecules, the ARO-MURI on Atomtronics, ARO MURI Quism program.

\bibliography{artikel,biblio1,footnotes}

\begin{thebibliography}{69}
\bibitem[{fno({\natexlab{a}})}]{fnot1}
\bibinfo{note}{In order that a true long range ordered state consistent with
  the Mermin-Wagner theorem is stable in the long time limit, a small hopping
  in the perpendicular direction is assumed.}
\bibitem[{fno({\natexlab{b}})}]{fnot2}
\bibinfo{note}{Numerical inaccuracies can lead to discrepancies between the
  result of the long-time non-equilibrium calculation and the expected final
  state equilibrium result.}
\end{thebibliography}

\end{document}